\documentclass[10pt,a4paper]{article}
\usepackage[margin=1in]{geometry}
\usepackage{graphicx}
\usepackage{hyperref}
\usepackage[utf8]{inputenc}
\usepackage[T1]{fontenc}
\usepackage{lmodern}
\usepackage{amsmath,amssymb,amsfonts}
\usepackage{amsthm}
\newtheorem{definition}{Definition}
\providecommand{\keywords}[1]{\par \vspace{1pc} \noindent \textbf{Keywords: }#1}
\date{}
\usepackage{booktabs}
\usepackage{tabularx}
\usepackage{array}
\newcolumntype{L}[1]{>{\raggedright\arraybackslash}p{#1}}
\newcolumntype{Y}{>{\raggedright\arraybackslash}X}
\usepackage{url}
\usepackage{xcolor}
\providecommand{\doi}[1]{doi:\href{https://doi.org/#1}{\nolinkurl{#1}}}

\hypersetup{
  colorlinks=true,
  linkcolor=black,
  citecolor=blue,
  urlcolor=blue,
  pdftitle={Ledger-State Stigmergy: A Formal Framework for Indirect Coordination Grounded in Distributed Ledger State},
  pdfauthor={Fernando Paredes Garc\'{i}a},
  pdfsubject={Neutral preprint on decentralized coordination over shared ledger state},
  pdfkeywords={Stigmergy, Distributed Coordination, Shared-State Coordination, Distributed Systems, Distributed Ledger, Multi-Agent Systems, Smart Contracts}
}

\newcommand{\St}{S_t}
\newcommand{\Stnext}{S_{t+1}}
\newcommand{\Oi}{\mathcal{O}_i}

\newcommand{\LSS}{\emph{ledger-state stigmergy}}

\title{Ledger-State Stigmergy: A Formal Framework for Indirect Coordination Grounded in Distributed Ledger State}

\author{
\bf Fernando Paredes Garc\'{i}a \\
Independent Researcher \\
\it fernando@develcuy.com
}

\begin{document}
\maketitle

\begin{abstract}
\noindent Autonomous software agents on blockchains solve distributed-coordination problems by reading shared ledger state instead of exchanging direct messages.
Liquidation keepers, arbitrage bots, and other autonomous on-chain agents watch balances, contract storage, and event logs; when conditions change, they act.
The ledger therefore functions as a replicated shared-state medium through which decentralized agents coordinate indirectly.
This form of indirect coordination mirrors what Grass\'{e} called \emph{stigmergy} in 1959: organisms coordinating through traces left in a shared environment, with no central plan.

Stigmergy has mature formalizations in swarm intelligence and multi-agent systems, and on-chain agents already behave stigmergically in practice, but no prior application-layer framework cleanly bridges the two.
We introduce \emph{Indirect coordination grounded in ledger state} (\emph{Coordinaci\'{o}n indirecta basada en el estado del registro contable}) as a ledger-specific applied definition that maps Grass\'{e}'s mechanism onto distributed ledger technology.
We operationalize this with a state-transition formalism, identify three recurring base on-chain coordination patterns (State-Flag, Event-Signal, Threshold-Trigger) together with a Commit-Reveal sequencing overlay, and work through a State-Flag task-board example to compare ledger-state coordination analytically with off-chain messaging and centralized orchestration.
The contribution is a reusable vocabulary, a ledger-specific formal mapping, and design guidance for decentralized coordination over replicated shared state at the application layer.
\end{abstract}


\section{Introduction}\label{sec:intro}

Blockchains started as a fix for the double-spending problem \cite{nakamoto2008bitcoin}.
After Ethereum added Turing-complete smart contracts \cite{buterin2014ethereum,wood2014yellow}, they turned into general-purpose shared-state machines: every participant reads the same world state, every transaction yields a deterministic state transition, and confirmed state is globally visible.
At the application layer, this creates a distributed-coordination problem over replicated shared state rather than over direct message exchange or centralized assignment.
An ecosystem of autonomous agents has grown up in this environment.
Arbitrage bots, liquidation keepers, maximal extractable value (MEV) searchers, and decentralized autonomous organization (DAO) delegates all coordinate without direct messages.
They read the chain, infer work from shared traces, and execute.

The coordination mechanism they rely on has a name, borrowed from entomology.
In 1959, Pierre-Paul Grass\'{e} noticed that termites build architecturally complex nests without blueprints or supervisors \cite{grasse1959}.
A termite drops a pheromone-laden mud pellet; that pellet stimulates the next termite to drop one nearby.
Grass\'{e} called the mechanism \emph{stigmergy}.
Swarm intelligence researchers adopted it \cite{bonabeau1999swarm,dorigo1996ant}, it showed up in robotic coordination \cite{parunak2005pheromones}, open-source development \cite{bolici2016stigmergy}, and Wikipedia editing \cite{elliott2006stigmergic}.
Heylighen \cite{heylighen2016stigmergy1,heylighen2016stigmergy2} generalized it into four components (agent, medium, trace, stimulation rule) and argued it is a universal coordination mechanism.

Application-layer connections between stigmergy and blockchain remain sparse.
G\"{u}rcan \cite{gurcan2022pow,gurcan2022aamas} modeled Proof-of-Work consensus as stigmergic, the most rigorous prior treatment, but his scope was limited to the consensus layer.
Capaccioli \cite{capaccioli2020blockchain} sketched a link between Bitcoin and swarm intelligence without operational definitions.
Dounas et al.\ \cite{dounas2022stigmergic} applied the concept to architectural collaboration on-chain but did not produce a general framework.
Adjacent swarm-robotics work has used blockchain smart contracts to coordinate physical or simulated robot swarms \cite{pacheco2020physicalswarm,pacheco2022foraging}, but its focus is robot meta-control rather than application-layer coordination patterns for software agents.
Existing work still stops short of mapping the full stigmergy apparatus onto application-layer ledger components and developing the resulting coordination patterns as a reusable catalogue for decentralized coordination over shared state.

This paper addresses that gap by treating the ledger as a coordination medium, not only as an execution substrate.

\medskip
\noindent\textbf{Definition.}
Building on Grass\'{e}'s notion of stigmergy (1959), we introduce \emph{Indirect coordination grounded in ledger state} (\emph{Coordinaci\'{o}n indirecta basada en el estado del registro contable}) as a ledger-specific applied definition of stigmergic coordination for distributed ledgers, where ledger state is the shared medium that agents observe and act upon.

\medskip
\noindent\textbf{Contributions.}

\begin{enumerate}
  \item We introduce \LSS{} as a named ledger-specific applied concept.
  The original stigmergy mechanism belongs to Grass\'{e} \cite{grasse1959}; the ledger-specific definition and application-layer specialization are introduced in this paper.
  \item We operationalize the definition with a state-transition formalism ($\St$, $\delta$, $\Oi$, $\mathcal{P}_i$) and identify three recurring base patterns together with a Commit-Reveal sequencing overlay for on-chain stigmergic coordination.
  \item We illustrate the framework through a State-Flag task-board example and use it to draw out the main trade-offs against off-chain messaging and centralized orchestration.
\end{enumerate}

\noindent A companion preprint \cite{paredesgarcia2026math} develops medium-agnostic mathematical foundations for stigmergic coordination over arbitrary shared media.
The present paper specializes that abstract framework to the distributed ledger domain and contributes the applied definition, design patterns, and application-layer analysis independently.
The formalism used here is deliberately lightweight: it provides a disciplined notation for ledger-observable traces and predicates, while the companion carries the theorem-level machinery.

\medskip
\noindent\textbf{Scope.}
We address application-layer \emph{coordination} over shared ledger state, not consensus mechanisms (covered by G\"{u}rcan \cite{gurcan2022pow}).
Token economics, governance design, and layer-2 scaling fall outside scope unless they directly affect the coordination patterns.
The paper argues through a running analytical example rather than a simulation study; limits appear in Section~\ref{sec:threats}.

\medskip
\noindent\textbf{Research Questions.}

\noindent\textbf{RQ1.} Can the classical stigmergy framework (medium, trace, agent, activation) be formally mapped onto distributed ledger components?

\noindent\textbf{RQ2.} What distinct on-chain coordination patterns emerge from treating ledger state as a stigmergic medium?

\noindent\textbf{RQ3.} What trade-offs emerge when shared-state stigmergic coordination is contrasted analytically with off-chain messaging and centralized orchestration?

\medskip
\noindent\textbf{Organization.}
Section~\ref{sec:background} reviews background.
Section~\ref{sec:model} defines the formal model.
Section~\ref{sec:patterns} develops the pattern catalogue.
Section~\ref{sec:casestudy} develops the running example.
Section~\ref{sec:evaluation} compares the coordination styles analytically.
Section~\ref{sec:discussion} discusses implications and limits.
Section~\ref{sec:related} covers related work, and Section~\ref{sec:conclusion} concludes.

\section{Background}\label{sec:background}

Three traditions converge in our definition: biological and computational stigmergy, environment-mediated coordination in multi-agent systems, and blockchains as state machines.

\subsection{Stigmergy: From Termites to Software}

Grass\'{e} \cite{grasse1959} coined \emph{stigmergie} (Greek \emph{stigma}, ``sign,'' plus \emph{ergon}, ``work'') from fieldwork on termite construction.
A termite deposits a mud pellet laced with pheromone; the pellet's presence stimulates a neighbor to deposit another nearby.
No termite carries a plan.
Structure emerges from local reactions to a shared physical environment.
Theraulaz and Bonabeau \cite{theraulaz1999history} refined the picture, distinguishing \emph{sematectonic} stigmergy (where the structure itself is the stimulus, as when a half-built column invites completion) from \emph{marker-based} stigmergy (where a separate chemical signal, like a pheromone trail, carries the information).

Dorigo et al.\ \cite{dorigo1996ant,dorigo2004aco} translated marker-based stigmergy into Ant Colony Optimization, showing that artificial pheromone trails on a graph can solve combinatorial problems.
Bonabeau et al.\ \cite{bonabeau1999swarm} organized these results under the label of swarm intelligence.
Parunak et al.\ \cite{parunak2005pheromones} applied digital pheromone fields to unmanned vehicle coordination.

Heylighen \cite{heylighen2016stigmergy1} produced the most general theoretical account, decomposing stigmergy into four components: (i)~an \textbf{agent} that acts, (ii)~a \textbf{medium} in which traces persist, (iii)~a \textbf{trace} left by the action, and (iv)~a \textbf{stimulation rule} linking traces to subsequent actions.
A companion paper \cite{heylighen2016stigmergy2} classified varieties along two axes: quantitative versus qualitative traces, and sematectonic versus marker-based.
His point was that \emph{any} system where agents modify a shared environment and those modifications guide later behavior qualifies as stigmergic.
The ``environment'' can be a termite mound, a software repository, a wiki page, or, as we argue here, a distributed ledger.
This four-component decomposition provides the scaffold for our ledger mapping in Section~\ref{sec:model}.

\subsection{Indirect Coordination in Multi-Agent Systems}

Computer science reached similar ideas through a different path.
Gelernter's Linda \cite{gelernter1985linda,gelernter1992coordination} introduced \emph{tuple spaces}: shared associative memory where processes coordinate by writing, reading, and consuming data without knowing each other's identities.
The producer does not need to know who will consume its tuple, or when.
Malone and Crowston \cite{malone1994coordination} framed coordination more broadly as managing dependencies among activities, a definition that covers both message-passing and environment-mediated approaches.

Weyns et al.\ \cite{weyns2007environment} argued that treating the environment as an inert container for agents misses a source of coordination.
They proposed the environment as a first-class abstraction.
Omicini et al.\ \cite{omicini2004artifacts} pushed further with \emph{coordination artifacts}: objects embedded in the environment that encapsulate interaction rules and mediate how agents behave together, without requiring agents to know about each other.
Smart contracts do exactly this on a blockchain, but the connection has not, to our knowledge, been made explicit in the literature.

\subsection{Blockchain as Shared-State Machine}

A blockchain is a replicated state machine \cite{garay2015bitcoin}.
Ethereum's world state $\sigma$ maps every account to its balance, nonce, code, and storage \cite{wood2014yellow}.
A transaction $T$ produces a deterministic transition: $\sigma_{t+1} = \Upsilon(\sigma_t, T)$.
Four properties matter here: the resulting state is globally visible (any full node can read it), immutable once confirmed, deterministic, and cryptographically authenticated.

These properties make the ledger a good fit for a stigmergic medium: a shared environment where every action leaves a persistent, visible, verified trace that other agents can react to \cite{szabo1997formalizing}.
Smart contracts, in this reading, are coordination artifacts \cite{omicini2004artifacts} embedded in that medium.
A contract holds rules (its bytecode) and state (its storage), and agents interact with each other only through it.

A Uniswap pool makes this concrete.
The contract stores token reserves; a large swap pushes the on-chain price away from the external market; that price gap is a trace.
An arbitrage bot sees the gap (its predicate fires), submits a corrective swap (its action), and the reserves update (a new trace that may trigger further agents).
The liquidity provider and the arbitrageur never talk.
All coordination runs through the pool's storage.
The pool contract plays the same functional role as the termite mound in Grass\'{e}'s original account: a shared medium that accumulates traces and stimulates action.

\subsection{The Gap}

Table~\ref{tab:literature} maps the key sources to our definition.
Each thread is well-developed on its own, but the intersection has barely been explored.
G\"{u}rcan \cite{gurcan2022pow} produced the strongest prior connection; his scope was consensus, not coordination patterns.
Capaccioli \cite{capaccioli2020blockchain} gave a conceptual sketch with no operational definitions.
Dounas et al.\ \cite{dounas2022stigmergic} stayed within a single application domain.
Adjacent swarm-robotics work has used blockchain smart contracts to coordinate physical or simulated robot swarms \cite{pacheco2020physicalswarm,pacheco2022foraging}, but its focus is robot meta-control rather than application-layer smart-contract coordination patterns for software agents.
A general, formalized bridge for the application layer is still missing.

\begin{table}[ht]
\centering
\caption{Literature mapping: concept, source, and role in our definition}
\label{tab:literature}
\small
\begin{tabularx}{\textwidth}{L{3.0cm}L{3.2cm}Y}
\toprule
\textbf{Concept} & \textbf{Key Source(s)} & \textbf{Role in Our Definition} \\
\midrule
Original stigmergy & Grass\'{e} (1959) \cite{grasse1959} & Indirect coordination via environmental traces. \\
\addlinespace
Swarm intelligence & Dorigo et al.\ (1996) \cite{dorigo1996ant} and Bonabeau et al.\ (1999) \cite{bonabeau1999swarm} & Computational viability of stigmergic optimization and digital pheromone trails. \\
\addlinespace
Universal stigmergy & Heylighen (2016a, 2016b) \cite{heylighen2016stigmergy1,heylighen2016stigmergy2} & Four-component decomposition we map onto ledger primitives. \\
\addlinespace
Tuple spaces & Gelernter (1985) \cite{gelernter1985linda} & Shared associative memory as coordination substrate. \\
\addlinespace
Coordination artifacts & Omicini et al.\ (2004) \cite{omicini2004artifacts} and Weyns et al.\ (2007) \cite{weyns2007environment} & Environment-embedded coordination objects that support our smart-contract reading. \\
\addlinespace
Blockchain state machine & Wood (2014) \cite{wood2014yellow} and Garay et al.\ (2015) \cite{garay2015bitcoin} & Formal state-transition semantics for the medium. \\
\addlinespace
PoW as stigmergy & G\"{u}rcan (2022) \cite{gurcan2022pow} & Closest prior work; consensus-layer only. \\
\addlinespace
Blockchain robot swarms & Pacheco et al.\ (2020, 2022) \cite{pacheco2020physicalswarm,pacheco2022foraging} & Close adjacent work; robot-swarm meta-control rather than application-layer software coordination patterns. \\
\addlinespace
Bitcoin as swarm & Capaccioli (2020) \cite{capaccioli2020blockchain} & Informal link; no operational definitions. \\
\addlinespace
Stigmergic DAO & Dounas et al.\ (2022) \cite{dounas2022stigmergic} & Domain-specific (architecture); no general framework. \\
\addlinespace
MEV / frontrunning & Daian et al.\ (2020) \cite{daian2020flash} & MEV bots behave stigmergically. Also the main threat model. \\
\bottomrule
\end{tabularx}
\end{table}

\section{Definition and Formal Model}\label{sec:model}

\subsection{The Definition}

Building on Grass\'{e} \cite{grasse1959}, we define:

\begin{quote}
\textit{``Coordinaci\'{o}n indirecta basada en el estado del registro contable.''}\\[4pt]
\textnormal{(Indirect coordination grounded in ledger state.)}
\end{quote}

\noindent This definition introduces the ledger-specific specialization used in the present paper and connects it to the medium-agnostic shared-medium framework developed in \cite{paredesgarcia2026math}.
It makes three commitments: the coordination is \emph{indirect} (agents do not exchange messages), the substrate is the \emph{ledger state} (world state plus event logs), and agents' actions are \emph{grounded in} observations of that state, meaning the state causally triggers the action.

The Spanish formulation anchors the concept in the Latin American research tradition while the English translation keeps it accessible.

\subsection{Mapping stigmergy components to ledger primitives}\label{subsec:mapping}

Table~\ref{tab:mapping} shows the mapping from Heylighen's four components \cite{heylighen2016stigmergy1} to their ledger equivalents.
This is not a loose analogy.
Each component has a concrete counterpart in the ledger architecture, and the relationships between them (agent acts on medium, medium presents trace, trace triggers agent) are preserved structurally.

\begin{table}[ht]
\centering
\caption{Stigmergy $\rightarrow$ ledger mapping}
\label{tab:mapping}
\small
\begin{tabularx}{\textwidth}{L{2.1cm}L{4.0cm}Y}
\toprule
\textbf{Stigmergy} & \textbf{Heylighen (2016)} & \textbf{Ledger Counterpart} \\
\midrule
Medium & Shared environment where traces persist & Ledger state: world state $\sigma$ (storage, balances) + event logs $\mathcal{L}$ \\
\addlinespace
Trace & Persistent modification left by action & State transition $\delta(\St, a)$ producing $\Stnext$; emitted events $e \in \mathcal{L}$ \\
\addlinespace
Agent & Autonomous entity that acts and perceives & EOA-controlled bot, smart contract actor, keeper, MEV searcher \\
\addlinespace
Stimulation rule & Trace triggers subsequent action & Predicate $\mathcal{P}_i(\St) \rightarrow$ act: agent $i$ reads state, checks condition, submits tx \\
\bottomrule
\end{tabularx}
\end{table}

\subsection{Minimal formalization}

Let $\mathcal{S}$ be the set of valid ledger states, $\mathcal{A}$ the set of valid actions (transactions), and $\mathcal{I}$ the set of agents.

\begin{definition}[Ledger-State Stigmergy]
A \emph{ledger-state stigmergic system} is a tuple
\[
\langle \mathcal{S}, \mathcal{A}, \mathcal{I}, \{\mathcal{V}_i\}_{i \in \mathcal{I}}, \delta, \{\mathcal{O}_i\}_{i \in \mathcal{I}}, \{\mathcal{P}_i\}_{i \in \mathcal{I}} \rangle
\]
where:
\begin{itemize}
  \item $\St \in \mathcal{S}$ is the agent-observable ledger record at block height $t$, comprising confirmed world state together with the authenticated logs and receipts that agents can inspect after confirmation.
  \item $a \in \mathcal{A}$ is an action (transaction) executed by some agent $i \in \mathcal{I}$.
  \item $\delta: \mathcal{S} \times \mathcal{A} \rightarrow \mathcal{S}$ is the deterministic state-transition function: $\Stnext = \delta(\St, a)$.
  \item For each agent $i$, $\mathcal{V}_i$ is an agent-specific information set (local view), typically a projection of that ledger record onto the data visible to $i$.
  \item $\Oi: \mathcal{S} \rightarrow \mathcal{V}_i$ is the observation function for agent $i$, projecting the global ledger record to that local view.
  \item $\mathcal{P}_i: \mathcal{V}_i \rightarrow \{0,1\}$ is the activation predicate. When $\mathcal{P}_i(\Oi(\St)) = 1$, agent $i$ selects and submits an action $a_i$.
\end{itemize}
\end{definition}

Throughout the pattern catalogue below, we sometimes write $\mathcal{P}_i(\St)$ as shorthand for $\mathcal{P}_i(\Oi(\St))$ when the observation map is not the analytic focus.
Action selection itself is left outside the tuple: the companion's broader system class adds explicit response maps for that step, while this paper keeps only the activation condition needed for the application-layer patterns.
The total transition $\delta$ is likewise an idealization; deployed transactions can revert or lose races, and the wasted effort discussed in Section~\ref{sec:evaluation} comes from those failed or displaced attempts.

\noindent\textbf{The stigmergic loop.}
Agent~$j$ executes $a_j$, producing $\Stnext = \delta(\St, a_j)$.
Agent~$i$, who has no idea $j$ exists, observes $\Oi(\Stnext)$.
If $\mathcal{P}_i$ fires, $i$ submits $a_i$, yielding $S_{t+2} = \delta(\Stnext, a_i)$.
That new state may trigger yet another agent.
No message passes between $i$ and $j$; collective behavior emerges from shared observation alone.

This tuple is a ledger-specific specialization of the abstract shared-medium system class
\[
\mathfrak{S} = \langle \mathcal{M}, \mathcal{A}, \mathcal{I}, U, \{O_i\}, \{P_i\}, \{R_i\} \rangle
\]
developed in \cite{paredesgarcia2026math}, where additional structure---explicit response maps, admissibility conditions, and a decay operator for non-persistent media---is axiomatized and proved.

The observation function $\Oi$ allows heterogeneous views: a liquidation bot watches collateral ratios while a governance delegate watches vote counts.
We impose no algebraic structure on $\mathcal{V}_i$ beyond set membership; in this applied paper it is simply the codomain needed to express agent-specific observations.
The predicate $\mathcal{P}_i$ is deliberately left abstract; it can encode a simple threshold check or a learned policy from a reinforcement learning agent.
The model also says nothing about observation latency.
In practice, there is always a delay between a state transition being committed and an agent detecting it, but that delay belongs to the implementation, not the model. In deployed ledgers, transactions are also batched into blocks, so an agent may act on a state it observed earlier than the state in which its transaction eventually executes; the contention and frontrunning discussed below arise from that operational gap rather than from extra structure in the tuple.

\subsection{Relationship to Heylighen's taxonomy}

In Heylighen's classification \cite{heylighen2016stigmergy2}, \LSS{} is primarily \textbf{sematectonic}: the state structure itself (a contract's storage layout, a pool's reserve ratio) is the stimulus, not a detached marker.
The traces are \textbf{quantitative}: balances, counters, and timestamps carry graded signals rather than binary ones.

One difference from biological stigmergy stands out.
Pheromone trails evaporate; ledger state does not.
Biology has built-in garbage collection.
On-chain, a stale trace persists unless the contract includes explicit expiry logic.
We revisit this in Section~\ref{sec:discussion}.

\subsection{Conceptual architecture}

\begin{figure}[ht]
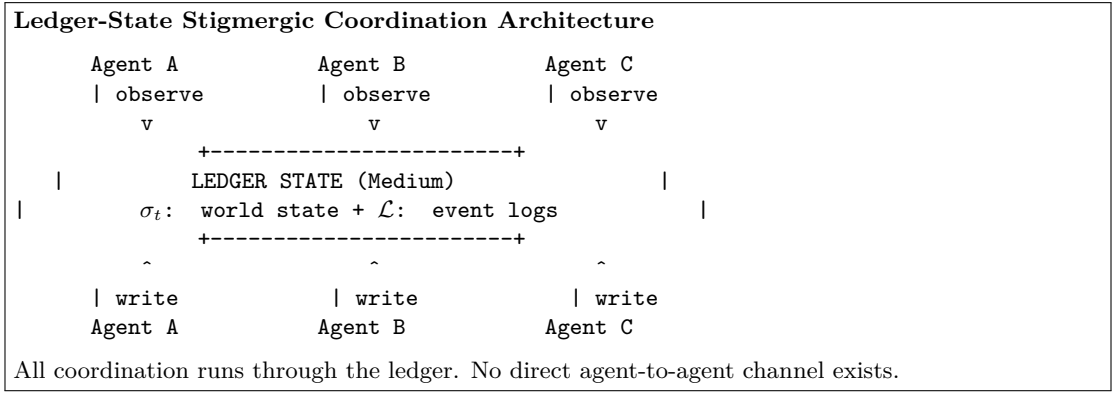

\centering
\fbox{\parbox{0.90\textwidth}{%
\small
\textbf{Ledger-State Stigmergic Coordination Architecture}\\[6pt]
\texttt{%
\begin{tabular}{@{}c@{}}
Agent A ~~~~~~~~~ Agent B ~~~~~~~~~ Agent C \\
~~|~observe~~~~~~~~~|~observe~~~~~~~~~|~observe \\
~~v~~~~~~~~~~~~~~~~~v~~~~~~~~~~~~~~~~~v \\
+-----------------------------------------------+ \\
|~~~~~~~~~ LEDGER STATE (Medium) ~~~~~~~~~~~~~~~| \\
|~~~~~~~~~$\sigma_t$: world state + $\mathcal{L}$: event logs~~~~~~~~~~~| \\
+-----------------------------------------------+ \\
~~\^{}~~~~~~~~~~~~~~~~~\^{}~~~~~~~~~~~~~~~~~\^{} \\
~~|~write~~~~~~~~~~~~|~write~~~~~~~~~~~~|~write \\
Agent A ~~~~~~~~~ Agent B ~~~~~~~~~ Agent C \\
\end{tabular}
}\\[6pt]
All coordination runs through the ledger.
No direct agent-to-agent channel exists.
}}
\caption{Conceptual architecture of ledger-state stigmergy}
\label{fig:architecture}
\end{figure}

Figure~\ref{fig:architecture} shows the basic layout.
Agents read the ledger (downward arrows) and write to it via transactions (upward arrows).
There is no lateral channel.

\section{Design Patterns for Ledger-State Stigmergy}\label{sec:patterns}

The general model admits different realizations depending on which part of the state carries the trace and how the activation predicate is structured.
We identify three recurring base patterns for single-contract ledger-state coordination together with one cross-cutting sequencing overlay, ordered by complexity.
They are best read as an applied pattern catalogue rather than as a closed classification of every possible ledger-state stigmergic design.

\subsection{Pattern 1: State-Flag}

A contract stores a Boolean or enumerated variable, say a task status that cycles through \texttt{OPEN}, \texttt{CLAIMED}, and \texttt{DONE}.
Agents poll it.
When the flag reads \texttt{OPEN} and the agent has the right capabilities, the agent claims the task.

The trace is a storage-slot update ($\texttt{flag}_{t} \neq \texttt{flag}_{t+1}$).
The activation predicate: $\mathcal{P}_i(\St) \equiv (\texttt{flag} = \texttt{OPEN}) \wedge (\texttt{agent}_i\texttt{.capable} = \texttt{true})$.

Gas cost is low (one \texttt{SLOAD} to read the flag) and the pattern is easy to audit.
The problem is contention.
If several agents see \texttt{OPEN} in the same block, they all submit a claim, and all but one waste gas.
Frontrunners can also snipe claims by watching the mempool and bidding higher gas prices \cite{daian2020flash}.
Combining State-Flag with commit-reveal (Pattern~4) or routing claims through a private mempool reduces this risk.
For griefing, where an agent claims a task and never finishes it, staking a bond with automatic reversion on timeout works well. Operationally, single occupancy in the worked contract comes from serial transaction execution together with a guard that causes later claimants to fail once the flag is no longer \texttt{OPEN}.
The companion preprint gives an abstract guarded-claim exclusivity result (Proposition~7.1 of \cite{paredesgarcia2026math}) that supports this single-occupancy intuition under serial execution.

Keeper networks like Gelato and Chainlink Keepers follow this pattern.
A contract exposes a machine-readable readiness check; keepers race to execute when that check indicates the task is actionable.

\subsection{Pattern 2: Event-Signal}

Rather than polling storage, agents subscribe to event logs.
A contract emits \texttt{TaskPosted(taskId, reward, deadline)} when a relevant state change occurs, and agents filter the log stream for matching events.

The trace is an event appended to $\mathcal{L}$.
Activation predicate: $\mathcal{P}_i(\St) \equiv \exists\, e \in \mathcal{L}_t : e.\texttt{type} = \texttt{TaskPosted} \wedge e.\texttt{reward} \geq \theta_i$.

Off-chain indexing (The Graph, WebSocket subscriptions) makes event monitoring cheaper than repeated \texttt{SLOAD} calls.
The catch is that events live in transaction receipts, not in the world state $\sigma$, so a smart contract cannot read its own past events.
This limits on-chain composability. Operationally, the pattern is hybrid: the triggering trace is on-chain, but most agents observe it through off-chain RPC or indexing infrastructure. What keeps the pattern inside this catalogue is the location of the coordination trace itself: the event is committed to the chain's authenticated record even when the observation path runs through middleware. The medium therefore remains ledger-grounded while the observation channel is operationally hybrid.
Spam is a risk too: a malicious contract can emit fake events to bait agents into wasted transactions.
Verifying the emitting address, cross-checking against storage state, and accounting for indexer lag or eclipse-style middleware failures are standard practice.
Reorganizations can cause events from orphaned blocks to vanish; waiting for $k$ confirmations before acting handles this.

Uniswap V2 and V3 emit \texttt{Swap} events that indexers and arbitrage bots consume.
Bots often subscribe to these events instead of polling reserves, since the logs summarize swap activity in one place.

\subsection{Pattern 3: Threshold-Trigger}

A quantitative state variable crosses a boundary: a lending pool's collateralization ratio drops below the liquidation line, a governance vote reaches quorum, or a token balance hits a cap.
Activation predicate: $\mathcal{P}_i(\St) \equiv \texttt{collateralRatio}(\St) < \theta_{\text{liquidation}}$.

Economic incentives fit naturally with this pattern: agents earn rewards for acting when thresholds are breached, and graded traces allow nuanced responses.
But Threshold-Trigger is the pattern most exposed to MEV.
Validators can see threshold-crossing transactions in the mempool and reorder them for profit \cite{daian2020flash}.
Batch auctions \cite{kelkar2020order} and MEV-share schemes offer partial mitigation.
Oracle manipulation is a second threat: if the threshold depends on an external price feed, a corrupt oracle can force false activations.
Time-weighted average prices and multi-oracle medians are the usual defenses.

Lending-protocol liquidations illustrate this directly. In Aave's terminology,
when a borrower's health factor drops below~1.0, any external agent can call the liquidation function and collect a bonus.
The health factor is the trace, the bonus is the incentive, and the liquidation call is the stimulated action.
No central party assigns liquidators.

\subsection{Pattern 4: Commit-Reveal Sequencing Overlay}

When frontrunning risk is high, one of the base patterns above can be wrapped in a two-round sequencing overlay.
Each agent first commits a hash of its intended action ($\texttt{commitments}[i] = H(a_i \| \texttt{salt}_i)$).
After a delay, agents reveal the plaintext.
The contract validates the hash and processes the action.

Phase~1 predicate: $\mathcal{P}_i^{\text{commit}}(\St) \equiv \texttt{phase} = \texttt{COMMIT} \wedge \neg\texttt{hasCommitted}(i)$.
Phase~2: $\mathcal{P}_i^{\text{reveal}}(\St) \equiv \texttt{phase} = \texttt{REVEAL} \wedge \texttt{hasCommitted}(i) \wedge \neg\texttt{hasRevealed}(i)$.

Direct mempool frontrunning is reduced substantially, but cost doubles (two transactions), latency increases by at least one block, and gas roughly doubles. Commitments also occupy storage until reveal or cleanup clears them.
The main risk is griefing by non-reveal: an agent commits then withholds the reveal.
Requiring a stake at commit time and slashing non-revealers addresses this, though it adds friction.
Commit-reveal is therefore best read as a cross-cutting sequencing overlay rather than as a fourth same-level trace family: it phase-gates actions over State-Flag, Event-Signal, or Threshold-Trigger structures when the value at stake justifies the overhead.

ENS domain auctions used this scheme historically.
Bidders committed hashed bids, then revealed them after a deadline.
Competitors could not see or outbid each other in real time.

\subsection{Pattern comparison}

Table~\ref{tab:patterns} summarizes trade-offs.

\begin{table}[ht]
\centering
\caption{Comparison of ledger-state stigmergy base patterns and sequencing overlay}
\label{tab:patterns}
\small
\begin{tabularx}{\textwidth}{L{2.0cm}Y Y L{1.1cm}L{1.25cm}Y}
\toprule
\textbf{Pattern} & \textbf{Signal} & \textbf{Activation} & \textbf{Gas} & \textbf{Delay} & \textbf{Main Threat} \\
\midrule
State-Flag & Storage var.\ & Bool/enum & Low & 1 block & Frontrunning \\
Event-Signal (hybrid) & Log entry & Subscription & Med. & 1 block+ & Spam, reorgs \\
Threshold & Numeric var.\ & Inequality & Med--High & 1 block & MEV, oracles \\
Commit-Reveal overlay & Two-phase & Phase-gated & High ($\sim$2$\times$) & $\geq$2 blocks & Non-reveal \\
\bottomrule
\end{tabularx}
\end{table}

The three base patterns are not mutually exclusive, and the Commit-Reveal overlay can wrap them when frontrunning risk justifies the added cost.
A lending platform might use Threshold-Trigger for liquidations and a Commit-Reveal overlay for governance votes within the same deployment.
We therefore treat these base patterns and overlay as an open analytical catalogue for single-contract coordination rather than as an exhaustive taxonomy. A further addition would need either a distinct trace/predicate structure or a distinct sequencing structure not reducible to these cases.

\section{Worked Example: On-Chain Task Board}\label{sec:casestudy}

To make the framework concrete, consider a \emph{decentralized task board}: a contract that posts tasks, and autonomous agents that claim and complete them using only the shared ledger state for coordination.
In the sense introduced in Section~\ref{sec:background}, the \texttt{TaskBoard} contract is the coordination artifact through which agents interact indirectly.
The example instantiates the State-Flag pattern and anchors the comparison developed below. It is intentionally a State-Flag witness rather than a worked demonstration of the full catalogue.

\subsection{Scenario}

A \texttt{TaskBoard} contract manages tasks $T_j$ with a reward $r_j$, a deadline $d_j$ (block number), and a difficulty indicator $w_j$ describing the effort needed to complete them.
Lifecycle follows State-Flag: $\texttt{OPEN} \rightarrow \texttt{CLAIMED} \rightarrow \texttt{DONE}$ (or $\texttt{EXPIRED}$).

Agents watch the board with heterogeneous capability profiles and finite operating budgets.
Their behavior is simple: read state, check for open tasks matching their capabilities, claim if one is available, and later submit completion.
This example is intentionally minimal so that the coordination mechanism remains visible instead of being buried under DeFi- or DAO-specific protocol detail.

\subsection{Three coordination styles}

\noindent\textbf{Off-chain messaging + on-chain execution (MSG).}
Agents discover tasks through a reliable off-chain pub/sub channel, negotiate claims off-chain, and then execute the agreed action on-chain.

\noindent\textbf{Centralized orchestrator + on-chain execution (ORCH).}
A trusted orchestrator assigns tasks by capability matching, informs agents off-chain, and lets them execute on-chain.
This is the efficiency-oriented baseline: it minimizes duplication of effort but introduces a single coordination authority.

\noindent\textbf{Ledger-state stigmergy (STIG).}
No off-chain coordination channel exists.
Agents observe only the \texttt{TaskBoard} state and act on their predicates. In deployment they may still rely on RPC or indexing infrastructure to observe that state, but that observation stack is not itself the coordination channel.
Trust rests solely on the ledger's consensus mechanism and on the contract's trace-management logic.

\section{Analytical Comparison of Coordination Styles}\label{sec:evaluation}

Table~\ref{tab:coordstyles} summarizes the comparison developed in this section. The strongest claims below are anchored in the State-Flag task-board witness; broader transfer to the rest of the catalogue remains a design hypothesis rather than an established general result.
Within that witness, the \texttt{TaskBoard} contract is the coordination artifact that internalizes the coordination rules into the shared medium instead of delegating them to a side channel or scheduler.

\begin{table}[ht]
\centering
\caption{Analytical comparison of coordination styles for the task-board witness}
\label{tab:coordstyles}
\small
\begin{tabularx}{\textwidth}{L{1.15cm}Y Y Y Y Y}
\toprule
\textbf{Style} & \textbf{Trust} & \textbf{Contention} & \textbf{Recovery locus} & \textbf{Transparency} & \textbf{Infra. dependency} \\
\midrule
STIG & Ledger consensus plus contract rules & High when many agents react to the same visible trace & On-chain medium: expiry, timeout, staking, slashing & High; traces and recovery logic are protocol-visible & Ledger plus practical RPC/indexing reliance for observation \\
MSG & Reliable off-chain coordination channel & Lower if negotiation succeeds before submission & Off-chain coordination and participant retry & Medium; agreement happens outside the ledger & Ledger plus messaging infrastructure \\
ORCH & Trusted scheduler or operator & Lowest in benign operation; scheduler bottlenecks remain & Central scheduler and its operational procedures & Low--medium; assignment logic is external & Ledger plus orchestrator service \\
\bottomrule
\end{tabularx}
\end{table}

\subsection{Benign trade-offs}

In benign conditions, the task-board witness makes ORCH the coordination-efficiency ceiling.
Because a single trusted scheduler assigns tasks, it avoids most duplicate claims and minimizes wasted action, though that efficiency comes with scheduler-side assignment overhead and a concentrated trust point.
MSG softens that advantage by distributing discovery and negotiation, but still relies on a side channel to keep collisions low before transactions reach the chain.

STIG removes that side channel entirely.
Its strength is trust minimization and public legibility: every agent can inspect the same contract state and decide locally whether to act.
It also shifts coordination liveness onto the ledger and the observation stack that agents use to follow its traces.
The price is contention.
If several agents see the same \texttt{OPEN} flag in the same block, they may all attempt to claim it, and all but one will waste effort or gas.
Within this State-Flag witness, that makes STIG the most transparent coordination style and usually the least coordination-efficient one in quiet conditions. We expect similar pressure wherever many agents are stimulated by the same visible trace, but this paper does not establish that as a general result for every pattern in the catalogue.

\subsection{Adversarial pressure}

The comparison changes once coordination logic must survive dishonest or unreliable participants.
A stigmergic design can embed recovery directly into the medium: staking, timeout reversion, claim caps, and explicit expiry all make the trace self-correcting without relying on a trusted scheduler.
That is the main architectural advantage of the approach.

However, this paper does not claim measured Byzantine superiority.
If STIG is equipped with timeout or slashing logic while ORCH and MSG are not, any advantage belongs to the full mitigation package, not to stigmergy in isolation.
A future empirical paper must test those baselines symmetrically before turning this analytical observation into a quantitative claim.

\subsection{Scaling considerations}

The central scaling variable is not merely the number of agents but the ratio of agents to simultaneously actionable traces.
A finite batch of open tasks invites races; continuous arrival or finer-grained partitioning would diffuse that pressure.
Under a ledger-state design, scalability therefore depends on how often many agents are stimulated by the same visible trace at once. That scaling story is clearest for collision-heavy State-Flag settings like the task board; transfer to Event-Signal, Threshold-Trigger, or overlay-heavy cases remains a hypothesis for later validation.

This leads to a practical design rule: reduce unnecessary trace collisions.
Partition work where possible, expose capability-relevant filters, and make stale traces expire or decay.
Those steps do not eliminate contention, but they keep a stigmergic system from degenerating into permanent claim races.

\subsection{Worked-example takeaway}

The task board highlights the main architectural trade-off cleanly.
MSG and ORCH externalize coordination into communication channels or scheduler-side assignment machinery.
STIG internalizes coordination through the contract-as-artifact together with the shared ledger medium.
That makes STIG more autonomous and more legible at the protocol boundary, but it also forces the contract designer to pay for contention management, stale-trace cleanup, and failure recovery explicitly.

\section{Discussion}\label{sec:discussion}

\subsection{When this approach makes sense}

On-chain stigmergic coordination costs more gas than stronger off-chain coordination mechanisms and introduces contention.
It is not always the right choice.
It pays off when trust assumptions are minimal, when there is no reliable off-chain channel, and when protocol-level legibility matters more than raw efficiency.
A rough heuristic is simple: if the cost of a coordination failure (a missed liquidation, a stuck task, an uncounted vote) exceeds the premium of keeping the coordination trace on-chain, the stigmergic approach is worth the extra machinery.
Otherwise, MSG or ORCH will often be cheaper and faster.

\subsection{Contract design lessons}

If autonomous agents will interact with your contract, make readable state a design priority.
Well-delimited status flags, monotonic counters, and documented storage layouts help agents evaluate predicates efficiently.
Staking, timeout reversion, and rate limits should be defaults for contracts meant for multi-agent use.

There is a specific design recommendation here: expose dedicated view functions that return what agents need to evaluate activation predicates.
A \texttt{getOpenTasks()} function is more agent-friendly than forcing agents to iterate a mapping and filter by status.
This amounts to ``stigmergic interface design'': making traces legible to the agents that read them.

\subsection{Security and incentive surface}

Public, legible traces are a double-edged design choice.
They let mutually distrustful agents coordinate without direct messaging, but they also expose the medium to frontrunning, spam, nuisance claims, and strategy leakage.
Rewards such as liquidation bonuses, keeper fees, or task payouts give agents reasons to react, yet the same visibility can make profitable traces copyable by faster or better connected actors.

That is why expiry, claim caps, staking, and timeout reversion are not just application logic but medium hygiene.
They bound state pollution and turn some harmful deviations into costly ones rather than free ones.
They also exploit a simple economic floor: flooding the medium with low-value traces is not free when the attacker must keep paying transaction fees to sustain the spam.
Where ordering games dominate, commit-reveal, batch execution, or private orderflow can reduce direct imitation, though each shifts assumptions or adds latency rather than eliminating the underlying stigmergic mechanism \cite{canidio2024commit,capponi2025mev,oz2024auctions}.
Privacy-preserving predicates and layered execution are natural extensions for the same reason: they preserve ledger-mediated coordination while changing what part of the trace is publicly legible and when it should be treated as final.

\subsection{Persistent versus decaying traces}

Pheromones evaporate.
Ledger state does not.
An expired task still marked \texttt{OPEN}, because nobody paid the gas to flip it, can trigger wasted agent actions indefinitely.
On-chain stigmergic state should include an expiration mechanism: a block-number deadline, an epoch counter, or an explicit invalidation call.

A more interesting approach is ``trace decay'' via decreasing rewards.
A contract could reduce the payout of a stale task over time, gradually weakening agents' predicates without requiring an explicit invalidation transaction.
Whether the gas cost of such logic is worth it remains context-dependent, but the design direction seems promising.

There is an analogy here to Dorigo's original ant colony optimization work \cite{dorigo1996ant}, where pheromone evaporation was not a bug but a feature.
Without evaporation, early pheromone trails would dominate forever, preventing the colony from adapting to changing conditions.
On-chain trace decay would serve the same function: letting stale coordination signals fade so that agents can redirect attention to current opportunities.
The difference is that on-chain decay costs gas, while biological evaporation is thermodynamically free.
The companion preprint \cite{paredesgarcia2026math} formalizes this distinction through a decay operator $D : \mathcal{M} \to \mathcal{M}$ applied between update steps, with persistent-trace systems recovered as the special case $D = \mathrm{id}$. More generally, forgetting response-relevant metadata such as timestamps collapses behavioral distinctions unless some replacement state structure is added back in.

\subsection{Cross-protocol stigmergy}

Our running example involved a single contract.
In practice, DeFi composability means that a trace in one contract routinely triggers actions in another.
A large deposit into a yield aggregator changes the aggregator's allocation strategy, which rebalances underlying lending pools, which shifts interest rates, which triggers borrowers to refinance.
Each step is a stigmergic reaction to a trace left by the previous step, and the chain of reactions can cross half a dozen protocols without any two participants being aware of each other.

This cross-protocol version of \LSS{} is more complex than what our model captures.
The observation function $\Oi$ would need to project over multiple contract states simultaneously, and the activation predicate might depend on relationships between states rather than simple thresholds within a single contract.
Formalizing this would require extending the tuple $\langle \mathcal{S}, \mathcal{A}, \mathcal{I}, \{\mathcal{V}_i\}_{i \in \mathcal{I}}, \delta, \{\mathcal{O}_i\}_{i \in \mathcal{I}}, \{\mathcal{P}_i\}_{i \in \mathcal{I}} \rangle$ to a multi-contract setting, likely with a product state space and cross-contract observation functions.
Such a product construction is developed abstractly in \cite{paredesgarcia2026math} (Proposition~7.2), where closure of admissible systems under product composition is proved; the companion also shows that freshness-sensitive behavior in a refined medium does not always admit a clean coarse-medium counterpart.
We leave this extension for future work, but note that the basic vocabulary (medium, trace, predicate) still applies; what changes is the topology of the observation graph.

\subsection{MEV as an adjacent broader case}

MEV searchers \cite{daian2020flash} are the most visible real-world instance of stigmergic competition around ledger systems, but they sit outside the paper's strict core object because they react not only to ledger state (world state plus logs) but also to pre-confirmation signals in the mempool and builder pipeline.
We therefore treat MEV as an adjacent broader case over a ledger-observable surface rather than as a pure single-medium instance of \LSS{}. Searchers watch pending order flow together with on-chain state, detect profitable configurations (arbitrage gaps, liquidatable positions), and submit transactions to capture them.

The analogy runs deeper than terminology.
Ants following pheromone trails can create traffic jams when too many converge on the same path.
MEV searchers competing for the same opportunity create gas-price bidding wars that congest the network.
Flashbots can be read as an attempt to channel this wild stigmergy into structured auctions, reducing the collateral damage.

\subsection{Limits of This Paper}\label{sec:threats}

\noindent\textbf{Analytical status.}
This paper does not report simulation outputs, deployment measurements, or quantitative benchmark data.
Its comparison among STIG, MSG, and ORCH is analytical: it identifies mechanism-level trade-offs and mitigation hooks without claiming measured performance gaps.

\noindent\textbf{Example scope.}
The task board is deliberately simple.
Real applications involve nested calls, cross-protocol dependencies, heterogeneous agent populations, and often continuous task arrival instead of a single fixed pool of actionable traces.
The example also instantiates only the State-Flag pattern, not the three base patterns plus the sequencing overlay.

\noindent\textbf{Fee and execution realism.}
A live network would add block-time variability, fee-market dynamics, builder incentives, mempool games, RPC/indexer latency, and richer adversarial behavior.
Those factors matter, but they are outside the evidence base of this paper and belong in a future empirical branch.
The same visibility that enables coordination can also leak agent intent and support spam or nuisance races; those risks are discussed qualitatively here but not measured.
Highly legible coordination for participating agents can also remain opaque to outsiders who do not share the same observation stack or strategic position.

\noindent\textbf{Future validation.}
The natural next step is not rhetorical expansion but empirical validation.
Any later paper that wants to claim quantitative robustness, gas bounds, or scaling behavior must supply a real simulator or deployment artifact together with recorded outputs.

\section{Related Work}\label{sec:related}

\noindent\textbf{Distributed coordination and shared-state systems.}
Our closest computer-science lineage outside blockchain-specific work is the literature on coordination through shared state and environment-mediated interaction.
Linda tuple spaces and coordination languages \cite{gelernter1985linda,gelernter1992coordination} showed how distributed processes can coordinate through shared associative memory without pairwise identity knowledge.
Malone and Crowston \cite{malone1994coordination} framed coordination more broadly as dependency management, while Weyns et al.\ \cite{weyns2007environment} and Omicini et al.\ \cite{omicini2004artifacts} treated the environment and coordination artifacts as first-class abstractions.
The present paper differs by specializing that lineage to replicated ledger state, where traces are globally visible, persistent, and authenticated by consensus rather than maintained by an external shared-memory service.

\noindent\textbf{Digital stigmergy.}
Parunak et al.\ \cite{parunak2005pheromones} applied digital pheromone fields to UAV coordination.
Bolici et al.\ \cite{bolici2016stigmergy} documented stigmergic patterns in open-source teams.
Elliott \cite{elliott2006stigmergic} argued that Wikipedia editing is stigmergic: editors modify a shared artifact and those modifications stimulate further edits.
Zheng et al.\ \cite{zheng2023wikipedia} provided quantitative evidence that edit traces predict subsequent editing activity.
None of these works address blockchain.

\noindent\textbf{Blockchain coordination.}
Lumineau et al.\ \cite{lumineau2021blockchain} studied blockchain governance as a coordination form but without the stigmergy lens.
Hassan and De Filippi \cite{hassan2021dao} formalized DAOs; whether DAO governance is better understood as stigmergic coordination than as voting remains an open question worth investigating.
Cong and He \cite{cong2019blockchain} analyzed how blockchain resolves information asymmetry in economic settings.

\noindent\textbf{Stigmergy $\times$ blockchain.}
G\"{u}rcan \cite{gurcan2022pow,gurcan2022aamas} is closest to us.
He modeled PoW as stigmergic consensus within a multi-agent systems paradigm: miners leave computational traces (valid blocks) that stimulate others to extend the longest chain.
We address the application layer instead of consensus, offer a general four-component mapping, and include an application-layer analysis.
Capaccioli \cite{capaccioli2020blockchain} linked Bitcoin to swarm intelligence in an informal essay.
Dounas et al.\ \cite{dounas2022stigmergic} applied stigmergy to on-chain architectural collaboration.
Adjacent swarm-robotics and decentralized-service work also treats shared ledger state as an indirect coordination substrate \cite{pacheco2020physicalswarm,pacheco2022foraging,vancalck2023market,queralta2023survey,liu2023seb}. That line of work is close in spirit, but its object is usually robot behavior, service orchestration, or application-specific cooperation rather than an application-layer coordination vocabulary for on-chain software agents.
These works do not provide the ledger-specific mapping, pattern catalogue, and analytical comparison developed here.

\noindent\textbf{Formal models and security.}
Garay et al.\ \cite{garay2015bitcoin} formalized the Bitcoin backbone protocol.
Luu et al.\ \cite{luu2016making} and Tolmach et al.\ \cite{tolmach2022survey} surveyed smart contract verification.
Castro and Liskov \cite{castro2002pbft} established PBFT.
Daian et al.\ \cite{daian2020flash} characterized MEV. Kelkar et al.\ \cite{kelkar2020order} proposed order-fair consensus.
Recent work also treats mitigation as part of the economic environment: Canidio and Danos \cite{canidio2024commit} analyze commitment mechanisms against front-running, Capponi et al.\ \cite{capponi2025mev} study private-orderflow adoption, and \"{O}z et al.\ \cite{oz2024auctions} document concentration in MEV-Boost block-building auctions.
We draw on this literature both as validation (MEV is stigmergic) and as threat model (stigmergic visibility is exploitable).

\noindent\textbf{Medium-agnostic foundations.}
Paredes Garc\'{i}a \cite{paredesgarcia2026math} develops an axiomatic framework for stigmergic coordination over arbitrary shared media, formulating a core system tuple, admissibility classes (persistent/decaying traces, exclusive-resource systems), response equivalence under observation-faithful representation, and a quotient adequacy/repair cycle for metadata refinements.
That work is deliberately medium-agnostic; the present paper provides the ledger-specific specialization, applied definition, design patterns, and application-layer analysis.

\section{Conclusion}\label{sec:conclusion}

We introduced \LSS{} as a lightweight formal framework for analyzing how autonomous agents coordinate through shared blockchain state, grounding the concept in Grass\'{e}'s 1959 mechanism \cite{grasse1959} and Heylighen's general framework \cite{heylighen2016stigmergy1}.
The four classical stigmergy components map onto ledger primitives, and from that mapping we identified an applied catalogue of three recurring base patterns---State-Flag, Event-Signal, and Threshold-Trigger---together with a Commit-Reveal sequencing overlay.

The State-Flag task-board example highlights the paper's central claim: trust-minimized coordination buys autonomy and protocol-level legibility at the price of extra contention, extra state-management logic, and reduced coordination efficiency relative to stronger off-chain coordination mechanisms.

Several things remain to be done.
We have not yet tested these patterns in a simulator or on a live network, where block timing, gas-price volatility, builder incentives, and adversarial adaptation would all matter.
Cross-protocol stigmergy, where a trace in one contract triggers action in another, is a natural next step; the medium-agnostic product composition of \cite{paredesgarcia2026math} provides an abstract basis, and DeFi composability offers a ready testing ground.
Formal verification of the patterns using established tools \cite{tolmach2022survey} would add confidence.
And the growing population of AI-powered on-chain agents will likely produce stigmergic dynamics that are more complex, and more consequential, than what we see today.


\end{document}